\documentclass[]{article}
\usepackage[]{geometry}
\usepackage{graphicx}
\usepackage{epsfig}
\usepackage{amssymb,amsfonts,amsmath,amsthm}
\usepackage{mathrsfs}
\usepackage{epstopdf}
\usepackage{hyperref,color}
\usepackage{natbib}
\usepackage{setspace}
\usepackage{enumitem}

\newtheorem{defi}{Definition}
\newtheorem{thm}{Theorem}
\newtheorem*{assum}{Basic Assumption}
\newtheorem{axiom}{Axiom}

\pdfoutput=1

\begin{document}

\title{Chapter 2: Spacetime from causality: causal set theory}
\author{Christian W\"uthrich and Nick Huggett\thanks{This is a chapter of the planned monograph \emph{Out of Nowhere: The Emergence of Spacetime in Quantum Theories of Gravity}, co-authored by Nick Huggett and Christian W\"uthrich and under contract with Oxford University Press. More information at \url{www.beyondspacetime.net}. The primary author of this chapter is Christian W\"uthrich (christian.wuthrich@unige.ch). This work was supported financially by the ACLS and the John Templeton Foundation (the views expressed are those of the authors not necessarily those of the sponsors). We wish to thank John Dougherty, Fay Dowker, David Meyer, David Rideout, and Sebastian Speitel for comments on earlier drafts.}}
\date{20 May 2020}
\maketitle

\label{ch:causets1}


\tableofcontents
\

\noindent
Causal set theory attempts to formulate a quantum theory of gravity by assuming that the fundamental structure is a discrete set of basal events partially ordered by causality. In other words, it extracts the causal structure that it takes to be essential for relativistic spacetimes, posits it as fundamental, imposes discreteness, and tries to establish that these spacetimes generically arise from the resulting structures in the continuum limit. Precursors can be found in David \citet{fin69}, Jan \citet{myr78}, and Gerard \citet{hooft79}, although the endeavour did not get started in earnest until 1987, when the seminal paper by Luca Bombelli, Joohan Lee, David Meyer, and Rafael Sorkin \citep{bomeal87} hit the scene. 

This chapter gives a brief introduction to the leading ideas of the program and offers a philosophical analysis of them. In \S\ref{sec:motivation}, we introduce the theory by putting it into its historical context of the tradition of causal theories of time and of spacetime and by showing how it grew out of concrete questions and results within that tradition. \S\ref{sec:kinematics} presents and discusses the basic kinematic axiom of causal set theory. Finally, in \S\ref{sec:space}, we will articulate what we will call the `problem of space' in causal set theory and illustrate the work required toward solving it. This will involve a consideration of what space is (\ref{ssec:essence}), how it might naturally be identified in causal set theory (\ref{ssec:spacecauset}), and how the dimension (\ref{ssec:dimension})), topology (\ref{ssec:topology}), and metrical properties such as distance (\ref{ssec:distance}) of such spatial structures could be determined.

\section{Motivation and background}
\label{sec:motivation}

\subsection{Historical prelude: causal theories of time and of spacetime}
\label{ssec:causaltheories}

Starting in 1911, just a short few years after Hermann Minkowski articulated the geometry of the spacetime of special relativity (SR), Alfred A Robb recognized that this geometry could be captured by the causal structure among the events of this spacetime. In fact, as \citet{rob14,rob36} proved, the causal structure of Minkowski spacetime determines its topological and metrical structure.\footnote{It can be shown that the group of all automorphisms of the causal structure of Minkowski spacetime is generated by the (inhomogeneous) Lorentz group---and dilatations, of course. This result was independently proved by \citet{zee64}, and apparently also by A D Aleksandrov in 1949. For a detailed analysis, cf.\ \citet{Winnie1977}.} In other words, the geometry of Minkowski spacetime can be fully reconstructed starting from the set of basal events and the binary relation of causal precedence in which they stand. As there are spacelike related events, there exist pairs of events such that none of them precedes the other. This signifies a loss of `comparability' and entails that the primitive causal relation imposes what mathematicians call a merely {\em partial} order on the set of events, as is appropriate for a special-relativistic theory.\footnote{Consult \citet[\S2.1]{hugeal12} for details. For a systematic account of the various attempts to axiomatize the structure of Minkowski spacetime and an assessment of the characteristics of the resulting logical systems, see \citet{Lefever2013}.} The spacelike relations between incomparable events can be defined in terms of combinations of fundamental causal relations, as there are always events in Minkowski spacetime which are to the causal future of any two spacelike related events. Thus, two spacelike-related events thus stand in an indirect causal relation.

The derivation of the full geometry of Minkowski spacetime from a few axioms mostly relying on a set of primitive events partially ordered by a primitive binary relation and the early interpretation of this relation as {\em causal} has led philosophers to causal theories of spacetime more generally, reinvigorating a venerable tradition of causal theories of time dating back at least to Gottfried Wilhelm Leibniz. This rejuvenated tradition starts off in Hans \citet{reichenbach1924,reichenbach1928,reichenbach1956} and continues in Henryk \citet{Mehlberg1935,Mehlberg1937} and Adolf \citet{grunbaum1963,grunbaum1967} as a causal theory of {\em time order}, rather than of spacetime structure. 
Arguing from an empiricist vantage point, the goal was to explicate temporal relations in terms of their physical, i.e., more directly empirically accessible, basis. In this spirit, Reichenbach postulates a set of events, merely structured by basic relations of a causal nature, where the relations of `genidentity' and of causal connection play a central role.\footnote{Two events are {\em genidentical} just in case they involve the same---i.e., numerically identical---object. It is thus clear that genidentity also involves, apart from mereological considerations, a causal connection between the genidentical events.} Reichenbach's early attempts to execute this program failed on  grounds of circularity:\footnote{Speaking of circularity, he rules out closed causal chains {\em ab initio}---on empirical rather than logical grounds.} his theory makes ineliminable use of the asymmetry of causal connection to ground the asymmetry of temporal order, but his criterion to distinguish between cause and effect---often called the `mark method'---relies, implicitly, on temporal order.\footnote{As criticized by \citet{Mehlberg1935,Mehlberg1937} and \citet{grunbaum1963}, among others.} The circularity arises because the distinguishability of cause and effect is necessary in an approach that assumes as fundamental an asymmetric relation and thus cannot take recourse to a more fundamental description of the events that may ground the distinction. Reichenbach's early formulation of the causal theory of time order also uses spatiotemporal coincidence as primitive and thus fails to explicate temporal order entirely in non-spatiotemporal terms.\footnote{Cf.\ \citet[\S6.2.a]{vanfraassen1970}.} His later formulation \citep{reichenbach1956} mends some of these deficiencies by attempting to explicate the causal asymmetry in terms of factual asymmetries in actual series of events ordered by temporal betweenness. This move does not get entirely rid of primitive temporal notions, but at least it provides an independent, `physical', grounding of the asymmetry of the causal connection. \citet{grunbaum1963,grunbaum1967} adopts Reichenbach's basic strategy, but tries to overcome the difficulties that befell its predecessor. The result is mixed, and no full explication of spacetime in terms of physical relations is achieved.\footnote{Cf.\ also \citet[\S6.3]{vanfraassen1970}.}

Bas \citet[Ch.~6]{vanfraassen1970} offers what he argues is a significant simplification of the theory that does not rely on purely spatial or spatiotemporal notions (182). His causal theory of {\em spacetime} (as opposed to just {\em time}) takes as primitive the notion of an {\em event} and the binary relations of {\em genidentity} and {\em causal connectibility}. Genidentity is an equivalence relation and is used to define world lines of objects as equivalence classes under genidentity, while the reflexive and symmetric relation of causal connectibility captures the causal structure of spacetime. Van Fraassen's final version of the theory---the details of which we will leave aside---dispenses with genidentity in favour of primitive persisting objects. 

Evidently, at least in the context of SR, the (possibly directed) causal connectibility relation is equivalent to the spatiotemporal relation of `spatiotemporal coincidence or timelike relatedness' (the first disjunct is necessary because of the reflexive convention chosen). This raises the worry---to which we will return below---whether the theory is trivial and the reconstruction hence pointless in the sense that it just restates the spatiotemporal structure of Minkowski spacetime in different terms and thus fails to ground that structure in more fundamental, {\em non-spatiotemporal}, relations. Against this charge, van Fraassen, like his precursors before him, insists that the causal theory outstrips the relational theory of spacetime because, unlike blandly spatiotemporal relations, the fundamental causal relations are {\em physical} relations. This line of defence may incite the further worry that the causal theory attempts to illuminate the obscure with the truly impervious: analyses of causation are famously fraught with a tangle of apparently impenetrable problems. This concern is allayed, van Fraassen retorts, by the fact that no general definition of either causation or physical relations is required for the project to succeed; since the non-modal component of `causal connectibility'---causal connection---gets explicated in terms of genidentity and `signal connection', and since these are evidently physical and empirical relations, `causally connectible' has a meaning which is not derivatively spatiotemporal. Furthermore, \citet[195]{vanfraassen1970} notes, the causal theory does not insist that all spatiotemporal relations are in fact {\em reduced}, but merely that they are {\em reducible}, to causal connectibility. We will need to return to this point below when we consider whether fundamental structures according to causal set theory are spatiotemporal at all. 

\subsection{Earman's criticism of causal theories of spacetime}
\label{ssec:earman}

Shortly after van Fraassen published what was then the most sophisticated articulation of the causal theory of spacetime, John \citet{Earman1972} aired devastating criticisms of the theory. Earman's objections traded on results in general relativity (GR) and established that no causal theory of spacetime has the resources to deal with the full range of spacetime structures licensed by GR. 

In order to appreciate Earman's points, and the story unfolding after his attack leading directly to causal set theory, a few concepts need to be introduced. A {\em spacetime} is an ordered pair $\langle \mathcal{M}, g_{ab}\rangle$ consisting of a four-dimensional Lorentzian manifold $\mathcal{M}$ with a metric field $g_{ab}$ defined everywhere on it. A {\em model of GR}, or just a `relativistic spacetime model', then is a triple $\langle \mathcal{M}, g_{ab}, T_{ab}\rangle$ including a stress-energy tensor $T_{ab}$ such that Einstein's field equation (EFE),
\begin{equation}\label{eq:einstein}
R_{ab} - \frac{1}{2} R g_{ab} + \Lambda g_{ab} = 8\pi T_{ab},
\end{equation}
given here in natural units $c=G=1$, where $R_{ab}$ is the Ricci tensor, $R$ the Ricci scalar, and $\Lambda$ the cosmological constant, is satisfied. A spacetime is {\em temporally orientable} just in case a continuous choice of the future half of the light cone (as against the past half) can be made across $\mathcal{M}$. Since a temporally orientable spacetime always affords a smooth, everywhere non-vanishing timelike vector field $t^a$ on $\mathcal{M}$,\footnote{\citet[Lemma 8.1.1]{Wald1984}.} we can take such a vector field to encode the temporal orientation of spacetime. In what follows, we will assume that all spacetimes are temporally orientable. If in fact the orientation is given, e.g.\ by a smooth timelike vector field, then we will call the spacetime temporally orient{\em ed}.

In order to get to the relevant causal structure, we need the notion of a timelike relation:
\begin{defi}[Timelike relation]\label{def:timelike}
Let $\langle \mathcal{M}, g_{ab}\rangle$ be a temporally oriented relativistic spacetime model. The binary relation $\ll$ of timelike separation is then defined as follows: $\forall p, q \in \mathcal{M}, p\ll q$ if and only if there is a smooth, future-directed timelike curve that runs from $p$ to $q$.
\end{defi}
The relation $\ll$, which is technically a relation of {\em timelike} separation in GR, will nevertheless be used to encode the {\em causal} structure. Let us grant, at least for the time being, that this relation is indeed fundamentally causal. We will return to this point below. 

Since we are interested in the causal {\em structure}, we need a criterion to identify causal structures. Here is the relevant isomorphism:
\begin{defi}[$\ll$-isomorphism]\label{def:ll-iso}
Let $\langle \mathcal{M}, g_{ab}\rangle$ and $\langle \mathcal{M}', g'_{ab}\rangle$ be temporally oriented relativistic spacetime models. A bijection $\varphi: \mathcal{M} \rightarrow \mathcal{M}'$ is a {\em $\ll$-isomorphism} if, for all $p, q \in \mathcal{M},
p\ll q$ if and only if $\varphi (p) \ll \varphi (q).$
\end{defi}
It is clear that two spacetimes have the same causal structure just in case there is a $\ll$-isomorphism between their manifolds. The causal structure is precisely what is preserved under these isomorphisms. Analogously, we can introduce the notion of a causal relation, which is only slightly more general than that of a timelike relation:
\begin{defi}[Causal relation]\label{def:causal}
Let $\langle \mathcal{M}, g_{ab}\rangle$ be a temporally oriented relativistic spacetime model. The binary relation $<$ of causal separation is then defined as follows: $\forall p, q \in \mathcal{M}, p < q$ if and only if there is a smooth, future-directed timelike {\bf or null} curve that runs from $p$ to $q$.
\end{defi}
We have highlighted in bold font the relevant difference to Definition \ref{def:timelike}. Causal relations $<$ also give rise to a corresponding isomorphism, denoted `$<$-isomorphism', in full analogy to the $\ll$-isomorphism as defined in Definition \ref{def:ll-iso}. Note that the two relations $\ll$ and $<$ are generally not interdefinable.\footnote{More precisely, as \citet{Kronheimer1967} show, it generally takes two of the three relations, causal, chronological, and null, to (trivially) define the third. All three relations can be reconstructed from any one only under the imposition of appropriate restrictions.} But since their physical meaning is closely related, we will generally only state the results using $\ll$.

The general goal for a causal theorist must be to take causal structure as fundamental and show how this structure grounds everything else about the spacetime. In particular, we want to see that its metric structure is determined by the causal structure. Thus, the success, or at least viability, of the causal program gets measured, as it were, by the extent to which the causal structure of relativistic spacetimes determines their metric (and topological) structure. 

It turns out this success cannot be complete, as \citet{Earman1972} notes. Conceding, as he does, the causal connectibility relation to his explicit targets \citet{grunbaum1967} and \citet{vanfraassen1970}, Earman demonstrates that there are relativistic spacetimes for which there is no hope that the causal program succeeds, thus showing that the transition from a causal theory of time to a causal theory of relativistic spacetime is not trivial. To add insult to injury, Earman concludes that the doctrines of the causal theorist about spacetime ``do not seem very interesting or very plausible'' (75). However, subsequent work by David Malament and others and in causal set theory has clearly nullified {\em that} appearance, as should become apparent below. 

Earman starts out by precisifying the notion of causal connectibility. In order to do that, let us introduce the notions of the `chronological future' and `past':
\begin{defi}[Chronological future and past]\label{def:chrono}
For all points $p\in\mathcal{M}$, let $I^+(p)$ and $I^-(p)$ be the {\em chronological future} and the {\em chronological past}, respectively, as determined by:
\begin{eqnarray*}
I^+(p) &:=& \{q: p\ll q\},\\
I^-(p) &:=& \{q: q\ll p\}.
\end{eqnarray*}
\end{defi}
The {\em causal} future and past sets, denoted $J^+$ and $J^-$, respectively, can be defined in complete analogy to Definition \ref{def:chrono}, substituting throughout the causal relation $<$ for the timelike relation $\ll$. These notions permit the precise articulation (of one notion) of `causal connectibility' of two events in a given spacetime as the physical possibility of a causal signal of nonnull (affine) length to connect them:
\begin{defi}[Causal connectibility]
An event $q \in \mathcal{M}$ is {\em causally connectible} to another event $p \in \mathcal{M}$ just in case $q \in J^+(p) \cup J^-(p)$. 
\end{defi}
Earman then asks whether the causal theorist can supply a criterion of spatiotemporal coincidence that relies purely on the basal causal notions. Suppose we have a set of otherwise featureless `events' that partake in relations of causal connectibility but that do not, fundamentally, stand in spatiotemporal relations. Van Fraassen (1970, 184) then defines pairs of events to be {\em spatiotemporally coincident} just in case they are causally connectible to exactly the same events, i.e., for every event $r$, $r$ is causally connectible to the one if and only if it is causally connectible to the other. As Earman points out, this criterion is only adequate for spacetimes which satisfy the following condition: for every pair of points $p, q \in \mathcal{M}$, if $J^+(p) = J^+(q)$ and $J^-(p) = J^-(q)$, then $p=q$. Since GR permits many spacetimes which violate this condition,\footnote{This condition is crudely violated, e.g., in spacetimes containing closed timelike curves, which consist in numerically distinct points with nevertheless all have the same causal future and past. For a more subtle and interesting example for a past but not future distinguishing spacetime as defined in Definition \ref{def:dist}, see \citet[Fig.\ 37]{hawkingellis}.} a causal theorist could never hope to reduce spatiotemporal coincidence to causal connectibility for {\em all} general-relativistic spacetimes. This condition is very closely related to another one:
\begin{defi}[Future (past) distinguishing]\label{def:dist}
A spacetime $\langle \mathcal{M}, g_{ab}\rangle$ is {\em future distinguishing} iff, for all $p, q \in\mathcal{M}$,
\[
I^+(p) = I^+(q) \Rightarrow p=q
\]
(and similarly for {\em past distinguishing}).
\end{defi}
If a spacetime is both future and past distinguishing, then we simply call it {\em distinguishing}. As we will see momentarily, the conditions of future and past distinguishing mark an important limit of what sorts of spacetimes a causal theory of spacetime can hope to successfully capture. This is perhaps not too surprising given that a distinguishing spacetime cannot contain any closed timelike or null curves and that non-distinguishing spacetimes do at least contain non-spacelike curves which come arbitrarily close to being closed. Relying on causal structure as it does, the causal program remains incapable in cases where this structure is pathological. However,  pathological causal structures are certainly possible in notoriously permissive GR \citep{smewut11}. And for \citet[78]{Earman1972}, there are no good reasons to reject those spacetimes violating the condition as physically unreasonable, even though that is sometimes done, and not just by defenders of the causal program. He thus concludes that the causal theory of spacetime does not command the resources necessary to offer a novel basis ``for getting at the subtle and complex spatiotemporal relations which can obtain between events set in a relativistic space-time background'' \citep[79]{Earman1972}.

Some spacetimes violating Earman's condition are such that $\mathcal{M}$ can be covered by a family of non-intersecting spacelike hypersurfaces. These spacetimes are topologically closed `timewise' and every event in $\mathcal{M}$ contains closed timelike curves. In this case, it is always possible to find an embedding space which is locally like $\mathcal{M}$ but does not contain closed timelike curves. Given that such embedding spaces are causally well-behaved, the causal theorist might be tempted to use their causal structure to recover the spacetime structure of the spacetime. Earman notes that such an enterprise would be bound to fail, given that the original spacetime and the embedding space also differ in global properties: in the former, `time' is closed, while in the latter it is not. Thus, the causal structure of the embedding space could not possibly render the correct verdict about such important global properties of the spacetime at stake. But notice that the causal set theorist could respond to this by weakening the ambition to recovering merely the {\em local}, but not necessarily the {\em global}, spacetime structure. 

In fact, this strategy could be more widely applied to those non-distinguishing spacetimes which can be partitioned into distinguishing `patches' of spacetime. A patch $\langle \mathcal{S}\subseteq\mathcal{M}, g_{ab |\mathcal{S}}\rangle$ of a spacetime $\langle \mathcal{M}, g_{ab}\rangle$ would be distinguishing just in case it is with respect to the events in the patch, but not necessarily with respect to those outside of it. This covers many, though not all, spacetimes which violate Earman's condition. In this case, the causal theorist could claim to be able to recover the spatiotemporal structure of the patch from the patch's causal structure. In this sense, the program could then still be executed to fruition `locally'. Thus, what Earman identifies as a problem for the causal theorist could be turned into a strategy to deal with many of the spacetimes ruled out of bounds by Earman. Of course, this strategy would only be acceptable if accompanied by a commensurately attenuated ambition of the program. 

Earman's criticism, though sound, may thus not spell the end of the program. Thus, the causal theorist can surely respond to Earman's conclusion that ``taking events and their causal relations as a primitive basis is not sufficient for getting at the subtle and complex spatiotemporal relations which can obtain between events set in a relativistic space-time background'' (79) with a dose of healthy revisionism. The causal theorist is bound to fail if her pretension was to produce an alternative foundation for full GR, as Earman conclusively establishes; however, if the aspiration is instead to offer either a merely `localist' reconstruction of spacetimes in a sector or GR or else---and more relevantly for our present purposes---a {\em distinct}, and perhaps more fundamental, theory, Earman's objections can be circumnavigated.\footnote{For a detailed assessment as to what extent a causal theory can succeed in grounding GR, see \citet{Winnie1977}.} The latter, of course, is precisely what causal set theory aims to provide, as we will see shortly.

\subsection{Malament's theorem}
\label{ssec:malament}

That the situation is not hopeless for the causal theorist, and, more specifically, that the causal structure often provides a powerful ground for the entire geometry of a relativistic spacetime, was established by remarkable results of Stephen Hawking and collaborators \citep{haweal76} and David \citet{mal77}, building on results regarding `causal spaces' in the seminal paper of \citet{Kronheimer1967}. It is these results by Kronheimer and Penrose, Hawking and collaborators, and Malament that motivate the causal set theory program. They can be thought of as an extension of Robb's reconstructive project from SR to GR. Mathematically, the basic idea is that we take a point set with its topology, its differential structure, and its conformal structure to extract the relation $\ll$ defined on the point set. Then we throw away everything except the point set and the relation $\ll$ and try to recover the entire spacetime geometry. This is accomplished---to the extent to which it is accomplished---in terms of implicit definitions using invariances under the relevant group of mappings. Metaphysically, the idea is of course that, fundamentally, there only is the point set structured by $\ll$ and every other aspect of spacetime geometry ontologically depends on that structure. 

Using the definitions above, as summarized by \citet[1400]{mal77}, the relevant result of Hawking, King, and McCarthy is the following theorem:
\begin{thm}[\citealt*{haweal76}]
Let $\phi$ be a $\ll$-isomorphism between two temporally oriented spacetimes $\langle \mathcal{M}, g_{ab}\rangle$ and $\langle \mathcal{M}', g'_{ab}\rangle$. If $\phi$ is a homeomorphism, then it is a smooth conformal isometry. 
\end{thm}
Homeomorphisms, i.e., continuous mappings that also have a continuous inverse mapping, are maps between topological spaces that preserve topological properties. Furthermore,
\begin{defi}[Conformal isometry]
A $\ll$-isomorphism $\phi$ is a {\em conformal isometry} just in case it is a diffeomorphism and there exists a (non-vanishing) conformal factor $\Omega: \mathcal{M}' \rightarrow \mathbb{R}$ such that $\phi_\ast (g_{ab}) = \Omega^2 g_{ab}'$.
\end{defi}
Thus, Hawking and collaborators showed that the following conditional holds: if two temporally oriented spacetimes have the same topology and causal structure, then they have the same metric, up to a conformal factor. This leads to the natural question of just under what conditions do two temporally oriented spacetimes with the same causal structure have the same topology? Obviously, if we knew the answer then we could state under what conditions two temporally oriented spacetimes with the same causal structure have the same metric, up to a conformal factor. Malament's result gives a precise answer to this question: just in case they are distinguishing, i.e., just in case they are both future- and past-distinguishing. Here is the theorem:
\begin{thm}[\citealt{mal77}]\label{thm:mala}
Let $\phi$ be a $\ll$-isomorphism between two temporally oriented spacetimes $\langle \mathcal{M}, g_{ab}\rangle$ and $\langle \mathcal{M}', g'_{ab}\rangle$. If $\langle \mathcal{M}, g_{ab}\rangle$ and $\langle \mathcal{M}', g'_{ab}\rangle$ are distinguishing, then $\phi$ is a smooth conformal isometry.
\end{thm}
It is important to note that neither future- nor past-distinguishability alone are sufficient to clinch the consequent \citep[1402]{mal77}. The theorem thus establishes that, for a large class of spacetimes, causal isomorphisms also preserve the topological, differential, and conformal structure, and hence the metrical structure up to a conformal factor. In this sense, we can say that the causal structure of a relativistic spacetime in this large class `determines' its geometry up to a conformal factor. It is worth keeping in mind that this `determination' is not underwritten by explicit definitions of the geometrical structure of these spacetimes in terms of their causal structures;\footnote{Whether or not such explicit definitions can be given remains an open problem (David Malament, personal communication, 24 April 2013).} rather, the argument proceeds by showing that maps between spacetimes that leave the causal structure invariant, will also leave the geometrical structure invariant (again, up to a conformal factor). Even though this may be too thin a basis on which to lay grand claims of ontological dependence, the results are sufficiently suggestive to motivate the entire program of causal set theory. 

Before moving to causal set theory proper, let us make a few remarks regarding its immediate prehistory. Considerations concerning the relations between discrete and continuum spaces in philosophy, physics and mathematics certainly predate causal set theory---in philosophy, for instance, by a mere couple of millennia. At least since the advent of quantum physics have physicists played with the idea of a discrete spacetime, which were seen to be motivated by quantum theory.\footnote{E.g.\ \citet{Ambarzumian1930}.} It was recognized early on that a fundamentally discrete spacetime structure would create a tension with the invariance under continuous Lorentz transformations postulated by SR, and that solving this problem would be crucial to finding a quantum theory of gravity.\footnote{For the earliest attempt to reconcile Lorentz symmetry with a discrete spacetime that we know of, cf.\ \citet{Snyder1947}. We will return to this issue in the next chapter.
} An early clear statement of something that starts to resemble the causal set program has been given by David \citet{fin69}. Finkelstein argues that the macroscopic spacetime with its causal structure may arise as a continuum limit from a `causal quantum space'. He asserts in the abstract of the paper that ``[i]t is known that the entire geometry of many relativistic space-times can be summed up in two concepts, a space-time measure $\mu$ and a space-time causal or chronological order relation $C$, defining a causal measure space'' (1261). Unfortunately, no reference or proof is given in the text in support of this claim; in fact, the claim is essentially repeated, but with the qualification that this is true only of ``many'' spacetimes dropped:
\begin{quote}
The causal order $C$ determines the conformal structure of space-time, or nine of the ten components of the metric. The measure on spacetime fixes the tenth component. (1262)
\end{quote}
This is effectively Malament's theorem without the relevant qualification that the spacetimes be distinguishing (and so is a false proposition)! Similarly, in another pre-causal-set-theory paper, Jan \citet{myr78} uses Malament's result, published the previous year, as his vantage point without giving a citation, merely claiming that
\begin{quote}
[i]t is a well-known fact that in the standard, continuum space-time geometry the causal ordering alone contains enough information for reconstructing the metric, except for an undetermined local scale factor, which can be introduced for instance by means of the volume element $dV$. (4)
\end{quote}
Myrheim's project can clearly be considered a precursor to causal set theory. He argues that the partial causal ordering can just as naturally be imposed on a discrete set as on a continuous one, and that the discrete set has the advantage of automatically providing a natural scale, unlike the continuous space. This natural scale, as Bernhard \citet[135, 149]{rie68} observed, results from the fact that a discrete space, but not a continuous space, possesses an `intrinsic metric' given by {\em counting} the elements. The {\em number} of elementary parts of the discrete space determines the volume of any region of that space, which is of course not the case for continuous spaces. Myrheim's goal is to show how one can recover, in a statistical sense, not only the geometry, but also the vacuum EFE for the gravitational field from the discrete causal structure. While the project is suggestive, and the goal the same as in causal set theory, it does not succeed.\footnote{Although Myrheim is aware (cf.\ 3) that this is not in general true, he proceeds as if the causal ordering of any relativistic spacetime is globally well-defined; he also seems to assume that the local uniqueness of geodesics can be extended globally (6), which is not in general true. There are other oddities in his analysis, for instance when he takes coordinates to be more fundamental than the metric (8).}

Around the same time, Gerard \citet{hooft79}, in just a few short pages (338-344), sketches a quantum theory of gravity closely related to causal set theory. While the use of lattices in quantum field theory is generally considered a pretense introduced to simplify the mathematics or to control vicious infinities, 't Hooft takes the point of view that in the case of gravity, a kind of a lattice ``really does describe the physical situation accurately'' (338). He, too, uses the unqualified and hence false version of the statement that Malament proved in its qualified form (340); he, too, makes no reference to either \citet{haweal76} or \citet{mal77}. Starting out from a basic causal relation, for which he offers the temporal gloss as ``is a point-event earlier than'' (ibid.), and requiring transitivity, 't Hooft asserts four basic assumptions of the theory. First, the causal relation gives a partial ordering of events and defines a lattice, i.e., a discrete structure. Supposing that a continuum limit of this fundamental structure exists, he demands, secondly, that this lattice contains all the information to derive a curved Riemann space in this limit; that, thirdly, the existence of this limit constrains the details of the fundamental partially ordered structure; and that, finally, a curved Riemann space with the appropriate signature comprises all information on the entire history of the universe it represents, i.e., the entire spacetime. The second and fourth assumptions jointly entail that the fundamental structure contains all physically relevant information of the spacetime it describes. 't Hooft then sketches how the continuum limit could be taken and briefly mentions that a cosmological model could be gained by simply adding the demand that there exists an event that precedes all others. The sketch omits any dynamics, which 't Hooft admits should be added such that it approximates the usual Einstein dynamics in GR and adds that we should expect such an action to be highly non-local.

't Hooft then closes his speculations with the following words:
\begin{quote}
The above suggestions for a discrete gravity theory should not be taken for more than they are worth. The main message, and that is something I am certain of, is that it will not be sufficient to just improve our mathematical formalism of fields in a continuous Riemann space but that some more radical ideas are necessary and that totally new physics is to be expected in the region of the Planck length. (344)
\end{quote}
Such a radically new start is precisely what the program of causal set theory seeks to offer.

\section{The basic plot: kinematic causal set theory}
\label{sec:kinematics}
 
Causal set theory as we now know it starts with the foundational paper by Luca Bombelli, Joohan Lee, David Meyer, and Rafael Sorkin \citep{bomeal87}, who, unlike Myrheim and 't Hooft, explicitly build on \citet{haweal76}, \citet{mal77}, and \citet{Kronheimer1967}. Advocates of causal set theory often insist that their theory consists of three parts, like any other physical theory: kinematics, dynamics, and phenomenology. Even though many physical theories may conform to this partition, there is, of course, nothing sacrosanct about this trinity. As long as the theory clearly articulates what it considers the physical possibilities it licenses to be and how this explains or accounts for aspects of our experience, it may have as many parts as it likes. Be this as it may, we will split the discussion of the theory into kinematics and dynamics, and deal with the former in this section and the latter in the next chapter.

\subsection{The kinematic axiom}
\label{ssec:kinematics}

Causal set theory assumes a four-dimensional point of view and, impressed with Malament's result, attempts to formulate a quantum theory of gravity {\em ab initio}, i.e., not by means of a quantization of GR or a `general-relativization' of the quantum physics of the other three fundamental forces. The four-dimensional viewpoint regards spacetime as a fundamentally inseparable unity. It thus stands in opposition to canonical approaches which divide four-dimensional spacetime into three-dimensional spacelike slices totally ordered by one-dimensional time. Spacetime points are replaced in causal set theory with elementary `events'. These events, just as points in the spacetime manifold, have no intrinsic identity; instead, they only acquire their identity through the relations in which they stand to other such events.\footnote{The theory thus lends itself rather directly to a structuralist interpretation \citep{wut12}.} Consonant with Malament's result, and as reflected in its name, causal set theory takes the relations which sustain the identities of their relata to be {\em causal} relations. In a relativistic theory, causal relations order their relata merely partially \citep[\S2.1]{hugeal12}. Finally, the fundamental structure is assumed to be `atomic', and hence discrete. We are thus left with a fundamental structure of otherwise featureless elements partially ordered by a relation of causal precedence into a discrete structure:
\newtheorem{theorem}{Basic Assumption}
\begin{assum}
The fundamental structure is a discrete set of elementary events partially ordered by a relation of causal precedence. In short, it is a {\em causal set}.
\end{assum}
This assumption is too weak to articulate a sufficient condition delineating the models of the theory; rather, it should be regarded as a conceptually central necessary condition. More precisely, it can be stated as follows:
\begin{axiom}[Kinematic Axiom of Causal Set Theory]\label{ax:kinem}
The fundamental structure is a {\em causal set} $\mathcal{C}$, i.e.\ an ordered pair $\langle C, \preceq\rangle$ consisting of a set $C$ of elementary events and a relation, denoted by the infix $\preceq$, defined on $C$ satisfying the following conditions:
\begin{enumerate}
\item $\preceq$ induces a partial order on $C$, i.e., it is reflexive $(\forall x \in C, x \preceq x)$, antisymmetric $(\forall x, y \in C,$ if $x \preceq y$ and $y \preceq x$, then $x=y)$, and transitive $(\forall x, y, z \in C$, if $x \preceq y$ and $y \preceq z$, then $x \preceq z)$. 
\item $\preceq$ is locally finite:\footnote{This condition may perhaps be more adequately called `interval finiteness' \citep[19]{dri13}, where an {\em interval} $[x,z]$ in a causet $\mathcal{C}$ is a subcauset that contains all events $y\in C$ such that $x\preceq y\preceq z$. Intervals are sometimes also called `Alexandrov sets'.} $\forall x, z \in C, |\{y \in C | x \preceq y \preceq z\}| < \aleph_0$, where $|X|$ denotes the cardinality of the set $X$ (sometimes called its `size'). 
\item $C$ is countable.
\end{enumerate}
\end{axiom}
A causal set, or `causet', can be represented by a {\em directed acyclic graph},\footnote{At least if we gloss over the fact that acyclicity is often defined in a way that renders causets `cyclic' because their fundamental relation contains trivial `cycles' $x\preceq x$. If readers are bothered by our sloppiness here they are invited to substitute `noncircular' for `acyclic'.} i.e., a set of nodes (usually represented by dots) connected by directed edges (usually represented by arrows) without directed cycles (which are cycles in which each node is traversed without changing direction). A causet is determined by its directed acyclic graph. As is standard in causal set theory, we will often use so-called `Hasse diagrams', which are directed acyclic graphs of the transitive reduction of finite partially ordered sets.\footnote{The {\em transitive reduction} of a binary relation $R$ on a domain $X$ is the smallest relation $R'$ on $X$ with the same transitive closure as $R$. The {\em transitive closure} of a binary relation $R$ on $X$ is a smallest transitive relation $R'$ on $X$ that contains $R$. Effectively, the transitive reduction of a graph is the graph with the fewest edges but the same `reachability relations' as the original graph, or, in the case at hand, the transitive reduction of a causet is the causet with all those causal connections removed that are entailed by transitivity. At least in the finite case, the transitive reduction of a causet is unique.} In Hasse diagrams, the direction of the edges is encoded in the relative positions of the connected vertices, and not by arrows. The convention is that an element `earlier' in the partial order is always drawn below a `later' element. Thus, an event $x$ is drawn below a numerically distinct event $y$ and there is a line connecting the two just in case $x$ immediately precedes $y$. 
\begin{figure}
\centering
\epsfig{figure=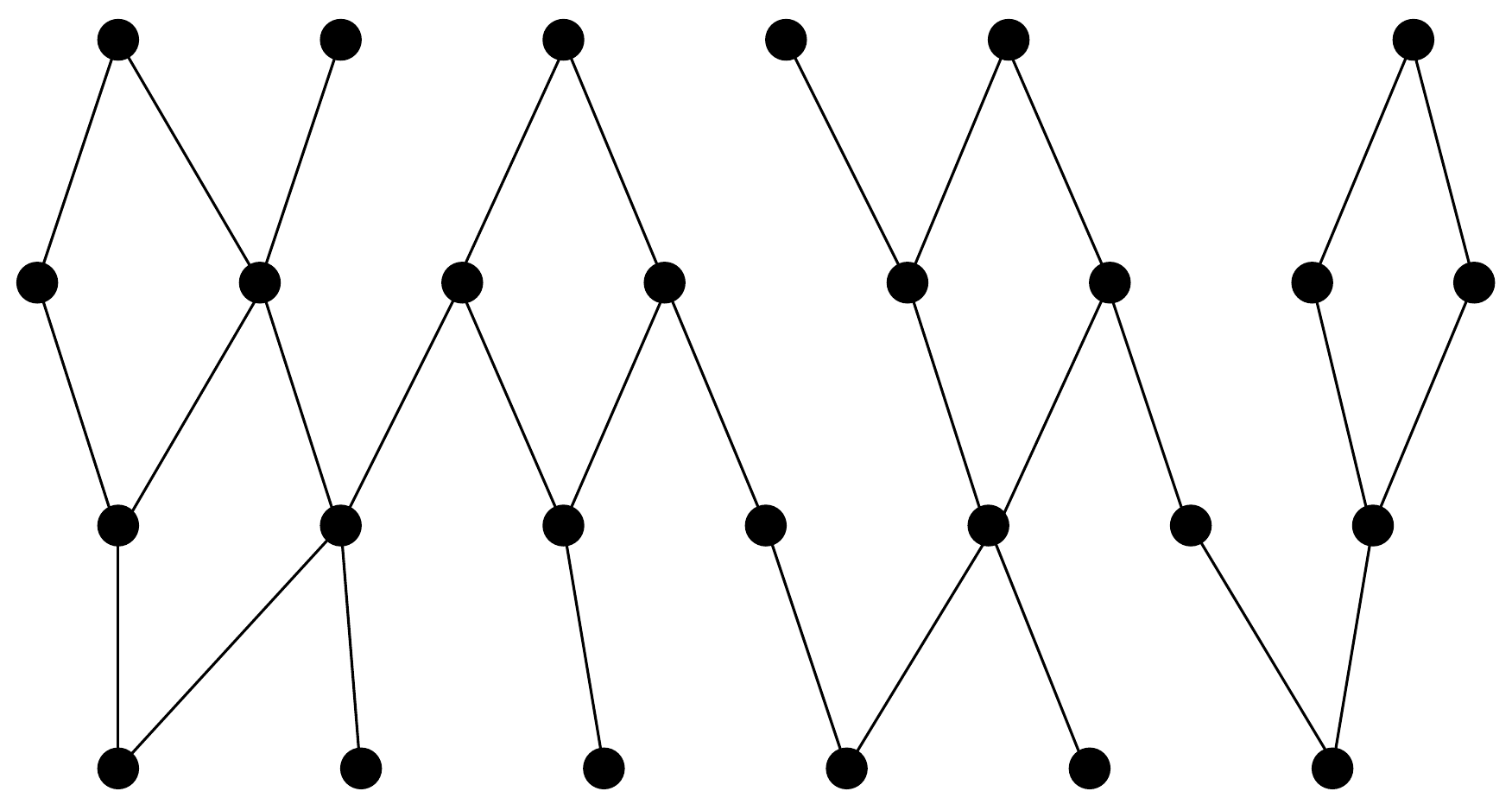,width=0.56\linewidth}
\caption{\label{fig:causet1} The Hasse diagram of a causet.}
\end{figure}
Figure \ref{fig:causet1} shows the Hasse diagram of a causet. Hasse diagrams are convenient because no two distinct causets can have the same Hasse diagram, so that a Hasse diagram uniquely determines the represented causet.\footnote{For a useful entry point onto the mathematics of partially ordered sets, cf.\ \citet{briwes00}. For a mathematically motivated and hence much wider perspective on causets, cf.\ \citet{dri13,dri17}.}

\subsection{Discussion of the basic assumptions}
\label{ssec:discuss}

A few remarks on the basic assumption and the kinematic axiom. First, it is important that the ordering be merely partial, and not total. A {\em total order} induced by a binary relation $R$ on a set $X$ is a partial order such that every pair of distinct elements of $X$ is {\em comparable}, i.e., $\forall x, y \in X$, either $Rxy$ or $Ryx$. That the ordering imposed by the causal structure is partial rather than total is crucial to capture the causal structure of a relativistic spacetime, where two distinct events can be spacelike separated and hence not stand in a relation of causal precedence. It is because the ordering is causal that it must be partial. 

Second, it is just as important that the ordering be no weaker than partial; in particular, it matters that it is not a mere pre-order. A binary relation on a set induces a {\em pre-order} just in case it is reflexive and transitive, but in general not antisymmetric. The additional imposition of antisymmetry rules out the possibility of causal loops of the form of cycles of distinct events $a, b\in C$ such that $a\preceq b \preceq a$. Two remarks concerning causal loops are in order. First, it should be noted that any events on such a loop would have the identical relational profile in the sense that any event that precedes one of them precedes all of them, and any event that is preceded by any one of them is preceded by all.\footnote{See \citet[236f]{wut12} for details.} If the relational profile of an elementary event constitutes its identity, as the rather natural structuralist interpretation of the fundamental structure in causal set theory would have it, events on a causal loop would therefore not be distinct at all and the theory lacks the resources to distinguish between a single event and a causal loop. This may be a limitation of the theory, but it reflects the assumption that any two events which do not differ in their causal profile do not differ physically and hence should be considered identical. Second, there are, of course, spacetimes in GR which contain causal loops \citep{smewut11}. Given that Axiom \ref{ax:kinem} prohibits causal loops at the fundamental level, it is not clear that causal set theory has the resources to show how relativistic spacetimes with causal loops can emerge from a fundamental causet without.\footnote{But see \citet{wut20}.} Many physicists consider causal loops `unphysical' and hence do not mourn their absence from the fundamental level (and so perhaps from emergent levels), but it is a cost that the theory incurs and that needs to be kept in mind. The fact that causal loops appear in so many relativistic spacetimes, including in (extensions of) physically important ones, should make us wary to dismiss them too precipitously, as is argued in \citet{smewut11} and elsewhere.\footnote{See the references therein. This caution certainly also motivates Earman's earlier criticism of the causal theory of spacetime: if all spacetimes violating Earman's condition are dismissed as unphysical, the causal theory survives his criticism unscathed.}

Third, the fundamental structure is assumed to be locally finite, and hence discrete. Unlike in loop quantum gravity, this discreteness is \emph{assumed} in causal set theory, not derived. A first set of justifications offered in the literature (e.g.\ in \citet[394]{hen09}) argue from the technical utility in assuming discreteness. A fundamentally discrete `spacetime' can cure a theory from divergences in various quantities that may not be tamed by renormalization, as well as simplify the computational challenges faces by the physicist. A second set of justifications typically given trades on physical reasons for preferring a discrete structure. For instance, without a short-distance cut-off as supplied by a discrete `spacetime', the semi-classical black hole entropy will not come out finite as desired (ibid.); or the discreteness may be an effective way of avoiding violations of the local conservation of energy in the form of photons with infinite energies \citep[6]{rei01}. Perhaps most curious of all, \citet[394]{hen09} argues that the concurrence of many rather diverse approaches to quantum gravity on the discreteness of the fundamental structure adds support to the stipulation of such discreteness in causal set theory. It is questionable, however, how different programs with contradictory assumptions which presuppose, or infer, the same discreteness can be considered mutually supportive. The truth of such an alternative theory would be evidence against causal set theory as a whole, though not necessarily against all its parts. Conversely, if causal set theory is true, or at least on the right track to a true quantum theory of gravity, as it presumably contends, then this entails that its competitors are mistaken. False theories may of course have true assumptions or implications; but whether this is so in any particular case can only be argued by a careful analysis of the relation between the distinct theories. Such an analysis is at best sketched in the literature; and of course, it may be that everyone is barking up the wrong tree.\footnote{See \citet[228f]{wut12} for a more detailed analysis of these arguments.}

Discreteness may give us other advantages. \citet[1f]{myr78} further points out that a fundamentally discrete `spacetime' carries an intrinsic volume and would thus give us a natural fundamental scale. This is again Riemann's point so revered by Myrheim's successors in causal set theory. As \citet[1]{myr78} also mentions, it may be that the discreteness has observable consequences. But to identify one particular assumption of an entire theoretical building as the one with a particular observable consequence is problematic; rather, a fully articulated theory is a whole package of assumptions, definitions, and techniques that jointly entail many consequences. Theory assessment, as most philosophers of science would insist, is typically an affair too holistic for us to be able to identify single assumptions engendering single consequences. Specifically, there are many ways in which a fundamentally discrete structure can be compatible with emergent continuous spacetime symmetries such as Lorentz symmetry (see below). Thus, it is far from obvious what the observable consequences of discreteness are, if any. Having said that, discreteness may well be a central axiom of an empirically successful theory and could thus be vindicated indirectly and a posteriori. With \citet{sor95}, one may simply consider fundamental discreteness necessary to ``express consistently the notion that topological fluctuations of \emph{finite} complexity can `average out' to produce an uncomplicated and smooth structure on larger scales'' (173). This is an appealing thought, though the apparent necessity may of course evaporate on closer inspection. In sum, although all the reasons given in favor of discreteness, as defensible as they are, are clearly defeasible, and, as we shall see in the next chapter
, there is a sense in which the discreteness leads to a form of non-locality, the proof, ultimately, is in the pudding: a theory's success ultimately validates its assumptions.

\begin{figure}
\centering
\epsfig{figure=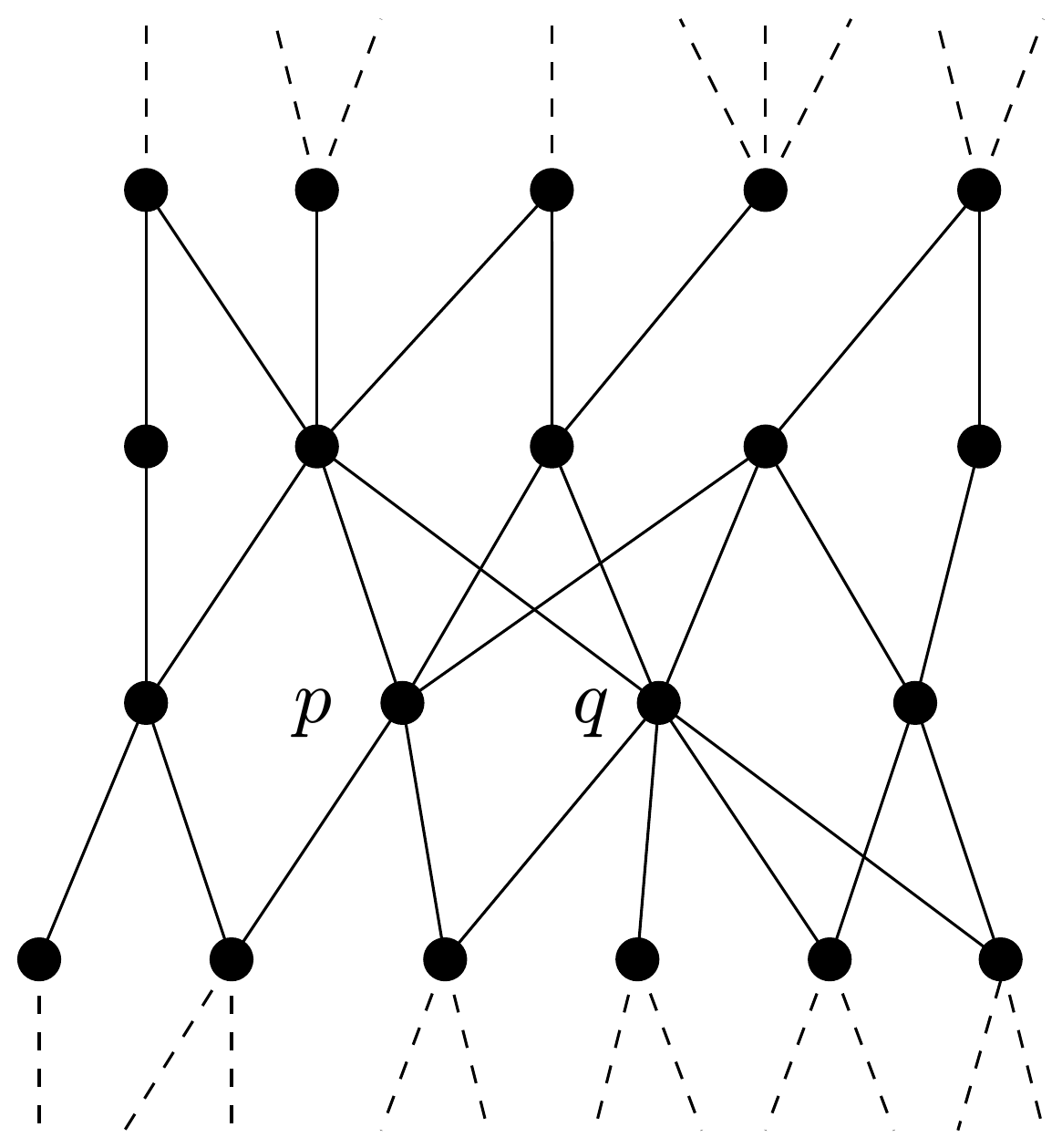,width=0.36\linewidth}
\caption{\label{fig:causet2} A past-distinguishing causet that fails to be future distinguishing.}
\end{figure}

Whatever its ultimate justification, it should be noted that the discreteness of causets has deep consequences. We will encounter some of these consequences over the course of this chapter and the next. One difference that matters in the present context arises if one considers whether an analogue of the program's basic motivating result---Malament's Theorem \ref{thm:mala}---holds in causal set theory. To state the obvious, this result is a theorem in {\em classical relativity theory}. There are other issues that would need to be settled for such a transliteration to be meaningful, but here it should be noted that an obvious causet analogue of future and past distinguishability is easily violated in causal set theory. Call a causet \emph{future (past) distinguishing} just in case for all pairs of events in the causet, if the set of events that are causally preceded by (causally precede) them are identical, then the events are identical. It is obvious that a causet such as the one depicted in Figure \ref{fig:causet2} violates this condition for the pair $p$ and $q$---both events have the same causal future.
That such a structure can satisfy the causal conditions while still violating distinguishability is a direct consequence of its discreteness. A causet which is not distinguishing and thus contains at least one pair of events with either an identical future or an identical past contravenes the premises of Malament's Theorem, but of course it is fully determined by its causal structure alone. It should be noted that in general a non-distinguishing causet still uniquely identifies each event in terms of its structure alone: the causet in Figure \ref{fig:causet2} is not future distinguishing, and hence not distinguishing, but the events $p$ and $q$ are structurally distinct in that they have different causal pasts. Thus, they fail to be a `non-Hegelian pair' \citep{wut12}. 

\subsection{Causation?}
\label{ssec:causation}

There remains an important point to be recorded before we move on. By virtue of what is the fundamental relation that structures the causets a \emph{causal} relation? One might object that such an austere relation cannot possibly earn this monicker because `causation'---whatever its precise metaphysics---refers to a much richer feature of our world. Why should one think that the partial ordering imposed on the fundamental and featureless events stands in any tangible connection to the causation that we attribute to the macroscopic and richly structured events that populate our world of experience? Of course, an answer to this question will probably require a solution to the measurement problem, and certainly a detailed understanding of how an effective spacetime emerges from underlying causets. We will turn to the latter problem in the next chapter
. But even granting---prematurely---a removal of these obstacles, there remain at least two worries. 

First, it may be argued that the conditions listed in Axiom \ref{ax:kinem} are not sufficient to make the relation at stake causal in the sense of relating causes and effects. The worry is familiar from relativity, and surely from the causal theory of (space)time, to which pedants (ourselves included) have always interjected that the `causal structure' of spacetimes merely captures a minimally necessary, but not sufficient, connection between events for them to be causally related as cause and effect. The objection is motivated by the observation that we do not attribute causal efficacy to all timelike or null relations; given an event, we take neither all events in its past lightcone to be its `causes', nor all events in its future lightcone to be its `effects'. But this interjection is neither very insightful, nor is it at all damaging to the causal set program. Those moved by it are hereby invited to mentally replace our talk of `causation' with more precise, but also more awkward, substitutions in terms of (possibly directed) `causal connectibility'.

Second, one may be worried that the conditions in Axiom \ref{ax:kinem} are not necessary for a relation to qualify as causal. In fact, all of the conditions---reflexitivity, antisymmetry, and transitivity---may be deemed unnecessary. Questioning the demand for reflexivity is rather straightforward, as on many accounts an event itself is not considered either among its own causes or effects. However, the demand for reflexivity is the least central to causal set theory. In fact, a new theory can be formulated that is `dual' to causal set theory except that the demand for reflexivity is replaced by one for irreflexivity. One sometimes finds the resulting irreflexive version of causal set theory, based on a strict partial order induced by an irreflexive relation $\prec$, in the literature.\footnote{A binary relation $R$ imposes a {\em strict partial order} on its domain $X$ just in case it is irreflexive (i.e., $\forall x\in X, \neg Rxx$) and transitive. A relation which induces a strict partial order is always anti-symmetric, i.e., irreflexivity and transitivity imply anti-symmetry. In fact, every such relation is asymmetric, i.e., irreflexivity and transitivity imply asymmetry. A binary relation $R$ is {\em asymmetric} just in case $\forall x, y \in X, Rxy \rightarrow \neg Ryx$.} If one presumes, quite reasonably, that whether all or none of the events stand in the fundamental relation with themselves can neither be directly observed nor does it have any empirical consequences, then the resulting dual theory is empirically equivalent to causal set theory as formulated above. In this sense, the choice between a reflexive and an irreflexive relation may be conventional.\footnote{Note that there is a bijection between `reflexive' and `irreflexive' causets: if $\preceq$ is a non-strict partial ordering relation, then the corresponding strict partial ordering $\prec$ is the {\em reflexive reduction} given by $x\prec y$ just in case $x\preceq y$ and $x \neq y$; conversely, if $\prec$ is a strict partial ordering relation, then the corresponding non-strict partial ordering relation is the {\em reflexive closure} given by $x\preceq y$ just in case either $x\prec y$ or $x=y$, for all events $x, y$ in the domain. The fundamental relations of the two dual theories, $\prec$ and $\preceq$, thus merely differ by the identity.} Thus, either a reflexive or irreflexive understanding of causation can arguably be reconciled with causal theory theory. What could not be accommodated, however, would be a non-reflexive notion of causation. Requiring causation to be antisymmetric may also be unnecessary, as an analysis of causation may want to leave room for cyclical causation as it can for instance be found in feedback loops. To this end, however, antisymmetry may only be given up at the level of event types, but not of event tokens. Finally, recent work on causation \citep[cf.~][Ch.~5]{paulhall} has found the requirement of transitivity unnecessary. 

Taking a step back from the details of these objections, the causal relations of causal set theory of course differs from that attributed to events in our ordinary lives. Nevertheless, given the tight relationship between the fundamental relation of causal set theory and the causal structure of relativistic spacetimes, we take it to be legitimate to dub this relation a `causal' relation.

\section{The problem of space}
\label{sec:space}

Having given and discussed the basic---kinematic---framework of causal set theory, a urgent question arises: what is the relationship between these causets and relativistic spacetimes? It cannot be one of identity. Mathematically, these structures are distinct; so distinct in fact that any physical interpretation of them which disregarded these differences would be in metaphysically negligent. But even if strict identity is not an option, the claim that the fundamental relation is causal could be rejected on the grounds that it still remains essentially spatiotemporal. This is a common objection to causal theories of (space)time, and is repeated in \citet{skl83}.\footnote{Sklar claims that on some versions of the causal theory of (space)time at least, the fundamental causal relations such as `genidentity' are based on what are at heart still spatio-temporal notions. Causal set theory does not, however, depend on the relations scrutinized by Sklar.} The objection is intended to lead to the conclusion that the reduction of (space)time structure to causal structure has thus failed; at best, what one gets is the elimination of some, but not all, spatiotemporal relations from the fundamental ideology. However, to interpret the causal relation in causal set theory as spatiotemporal would be disingenuous. The guiding idea of causal set theory is to offer a theory in which relativistic causation is fundamental and spacetime emerges. In that sense, all spatiotemporal structure is grounded in causal structure; spacetime ontologically depends on causal structure, but not vice versa. So let's take this idea seriously and see how far we can run with it.

In order to appreciate that the objection that the causal relation is just spatiotemporal has no purchase in causal set theory---unlike in GR---, let us consider just how different causal sets and spacetimes are. In the next chapter, we will see that causal sets are infested with a more virulent form of non-locality than we find in relativistic spacetimes. Furthermore, causal sets are discrete structures. This means that quite a bit of the geometric structure that we routinely attribute to space and time---and that is certainly available in GR---is simply missing in a causet. It is also important to note that the fundamental relation is {\em causal}, not {\em temporal}; temporal relations are supposed to emerge. Strictly speaking, thus, there is no time in causal set theory, or at least not at the kinematic level. Having said that, however, despite the intended difference, $\prec$ shares some properties with a generic relation of {\em temporal} precedence.\footnote{The intended relation is more generic than the relation of `chronological precedence' we find in the relativity literature, e.g.\ \citet{Kronheimer1967}; in fact, the latter relation is arguably essentially {\em causal}, rather than the other way around.} Just as with temporal precedence, we can define it as a reflexive or an irreflexive relation. If we further prohibit temporal loops and assume the transitivity of temporal precedence, both relations are relations are at least anti-symmetric (on the reflexive convention) or even asymmetric (on the irreflexive convention). Either way,  any relation of temporal precedence consistent with relativity, just as causal precedence, orders its domain only partially, not totally. The causal precedence relation at work in causal set theory does not pick a `Now' or gives rise to a `Flow'. But neither does temporal precedence on a B-theoretic metaphysics of time. In sum, then, the causal precedence of causal set theory though distinct has structural similarity with a stripped down B-theoretic version of (special-)relativistic time without metric relations such as durations.\footnote{\citet{dowker20} thinks that already the causal structure of relativistic spacetimes used as a vantage point for causal set theory really is a structure ``of \emph{precedence}, of \emph{before and after}, not of \emph{causation}'' (147n) and that causal set theory should thus have more appropriately been called `temporal set theory'. Unilluminating semantics aside, there is a clear sense in which the causal structure in GR is fundamentally and primarily \emph{causal}, and only derivatively temporal, as it encodes relations of causal connectibility of events by light signals. This means, to repeat, that the fundamental relation of causal set theory is best interpreted as `causal connectibility' or `causal precedence'.}

It should be noted that the remainder of the chapter concerns predominantly the presence (or absence, as the case may be) of spatial structures in causets in an attempt to grasp their connection to physical space. We will only be concerned with spacetime insofar as it may offer a path to space. 

\subsection{The `essence' of space}
\label{ssec:essence}

This leaves us somewhat inconclusive as regards time in {\em kinematic} causal set theory. It should be noted that we will encounter a much deeper problem, the so-called `problem of time' in canonical GR and quantum theories of gravity based on it in the chapters on loop quantum gravity. 
In the remainder of this section, we will argue that there is a similar `problem of space' in causal set theory, i.e., that `space' in particular is absent from fundamental causal set theory. The next chapter will return to time in the context of causal set theory and offer a fuller analysis, which will include its dynamics. 

In order to ascertain the absence of space, we need to state what kind of thing `space' is. Of course it is notoriously problematic to proclaim what the essence of `space'---or anything else for that matter---is, but we are going to do it anyway. Lest the readers mistake this for our being hubristic, 
let us hasten to reassure them that we do not take this list to express any kind of deep and final truth about the nature of space. We merely offer it as a useful starting point to determine the absence of space in fundamental causal set theory. No particular item on the list will be necessary for this purpose, they are all individually expendable; similarly, further items can be added to the list without threatening the purposes of this exercise---if anything, this would strengthen the case. The only thing that would undermine our argument for the absence of `space' would be if `space' had no nature at all---because that's precisely what we will find: `space' in causal set theory has no structure at all and hence no natural properties whatsoever. 

In his famed {\em La Science et l'Hypoth\`ese}, Henri \citet{poincare05} confidently asserts at least some essential properties of physical space:
\begin{quote}
In the first place, what are the properties of space properly so called?... The following are some of the more essential:---

1st, it is continuous; 2nd, it is infinite; 3rd, it is of three dimensions; 4th, it is homogeneous---that is to say, all its points are identical one with another; 5th, it is isotropic. (52)\footnote{Interestingly, the English translation omits the clause from the French original where the reader is informed that `isotropic' is to say that all straight lines through one point are identical with one another.}
\end{quote}
An attentive metaphysician will have a few quibbles with this passage. First, essentiality is not generally regarded as admitting of degrees. Second, Poincar\'e seems to employ either an awfully strong condition of homogeneity (and, in the original, of isotropy) when he paraphrases it as amounting to the identity of the points (or of all straight lines through a point), or else a notion of identity weaker than numerical identity. The former seems indefensible, as it would imply, among other things, that space cannot be extended---unless, of course, a point itself is extended. Points can be extended and cover entire `spaces', but only in non-Archimedean geometry \citep{ehr15}. Assuming standard geometry, homogeneity should be taken to require that all points of space share their essential physical properties but remain numerically distinct. 

Let us take Poincar\'e's lead though not his list and compile a list of essential properties of physical space, which seems better equipped to serve our purposes.\footnote{See \citet{HilbertHuggett} for a presentation of Poincar\'e's different purposes.} Space as we know and love it seems to have
\begin{itemize}
\item the structure of a differentiable manifold with a topology,
\item affine structure,
\item metric structure,
\item and a dimensionality.
\end{itemize}
`Space' as it is represented in various approaches to quantum gravity fails to have several or all of these features. In this sense, we have a `problem of space'! This problem is particularly pronounced in causal set theory. But first a few comments on the list. First, we do not mean to claim that `manifest' space must have these properties, but only that GR states space(time) does. So arguably, a precisification of `manifest' space would yield a theory that ascribes these properties, or properties very much like these, to space. Furthermore, although we will be concerned with spacetime as it figures in GR, the above list makes for perfectly natural assumptions about the nature of space in a much larger class of spacetime theories. 

\subsection{The space in a causal set}
\label{ssec:spacecauset}

So how could one identify `space' in a causet? What is the most natural conception of space `at a time' in causal set theory? Given some vantage point, i.e., some particular basal event, how can one determine the set of events that are `simultaneous' to it and jointly form `physical space' at an instant? There are obvious worries here concerning the relativity of simultaneity, but let's leave those to the side for the moment. In GR, at least in its globally hyperbolic sector, we can foliate the four-dimensional spacetime into totally ordered three-dimensional spacelike hypersurfaces that the metaphysician might consider identifying with `space at some time'. Of course, the choice of foliation of a relativistic spacetime is highly non-unique and in general not justifiable on physical grounds that would uniquely privilege the particular foliation chosen. Our best shot at a similar construction in causal set theory would be to partition the causal sets into totally ordered maximal sets of pairwise `spacelike' related events. Just as the foliation of globally hyperbolic spacetimes was non-unique, such partitions will in general not be unique in causal set theory. Furthermore, no two events in any such set can stand in the fundamental causal relation---if they did, one would causally precede the other and hence they could not be part of `space' at the same `time'. This means that the resulting subsets of events of the causet would, by necessity, be completely structureless---no two points in any subset could be related by the fundamental relation. Technically, this means that these subsets are `antichains':
\begin{defi}[Chain]\label{def:chain}
A {\em chain} $\gamma$ in a causet $\langle C, \preceq\rangle$ is a sequence of events in $C$ that are pairwise comparable, i.e., for any two events $x, y$ in $\gamma$, either $x\preceq y$ or $y\preceq x$. This implies that a chain is a subset of $C$ that is totally ordered by $\preceq$ (hence `sequence' rather than just `set').
\end{defi}
\begin{defi}[Antichain]\label{def:antichain}
An {\em antichain} $\alpha$ in a causet $\langle C, \preceq\rangle$ is a subset of events in $C$ that are pairwise incomparable, i.e., for any two events $x, y$ in $\alpha$, $\neg (x\preceq y)$ and $\neg (y\preceq x)$. This implies that an antichain is a subset of $C$ that remains completely unstructured by $\preceq$. 
\end{defi}
Given that they are altogether unstructured, an antichain is completely characterized by its cardinality, at least intrinsically. Extrinsically, it is characterized by how its elements are embedded in the total structure of the causet. The extrinsic characterization of antichains matters for our purposes, since it is important to {\em partition}, i.e., to divide into non-empty and non-overlapping subsets without remainder, the entire causet in order to obtain what can reasonably be considered a `foliation'. Furthermore, the elements of the partition---the antichains---should be {\em inextendible}, i.e., they should be such that any basal elements of the causet not in the antichain is related to an element of the antichain by $\preceq$. The problem, to repeat, is that antichains have no structure, and hence the fundamental causets have no (intrinsic) {\em spatial} structure at all, no metrical structure, no affine and differentiable structure---there is no manifold---, and---if at all---very different dimensionality and topology than the continuous spacetime they are supposed to give rise to. Furthermore, there is no evident spatial ordering, such as we ordinarily find in terms of spatial proximity. 

Before we have a closer look at some of these claims, let us state the important fact that at least for any finite partially ordered set, such a partition into antichains is always possible.\footnote{A partition of any causet into antichains is always trivially possible: just take the antichains each consisting of a single event. That {\em such} a partition would not satisfy our present needs should be clear, though. The point of the following is to establish that for a large class of causets, we are guaranteed that partitions conducive to these needs exist.} We need some definitions before we can state the theorem:
\begin{defi}[Height and Width]
The {\em height} of a partially ordered set $\mathcal{P}$ is the cardinality of the largest chain in $\mathcal{P}$. The {\em width} of $\mathcal{P}$ is the cardinality of the largest antichain in $\mathcal{P}$. 
\end{defi}
This allows us to formulate the theorem \citep[55]{bri97}:
\begin{thm}\label{thm:dilworth}
Let $\mathcal{P} = \langle P, \leq\rangle$ be a finite partially ordered set with height $h$ and width $w$. Then
\begin{enumerate}[label=(\alph*)]
\item the domain $P$ can be partitioned into $h$ antichains;
\item the domain $P$ can be partitioned into $w$ chains.
\end{enumerate}
\end{thm}
As a corollary, this gives us an upper bound for the cardinality of $P$: $|P| \leq hw$. The part (b) is a major result and requires a much more sophisticated proof and is also known as `Dilworth's Theorem' (ibid.). For our purposes, however, the first part (a) suffices to establish that for a finite causet, a partition into a sequence of `Nows' is always possible. That the result can straightforwardly be generalized to past-finite causets---a class of causets that will take center stage in the next chapter---is clear from the proof of (a) \citep[56]{bri97}. This proof proceeds by constructing antichains as follows. Define the height $h(x)$ of an event $x$ in $P$ as the cardinality of the longest chain in $\mathcal{P}$ with top element $x$, i.e.\ the maximal element in the chain. Then collect all the events of the same height into a set, i.e., a set of events of height 1, a set of events with height 2, etc. Every event has a height and will thus be an element of one of these sets. These sets turn out to be antichains, which completes the proof. Now the extension of Theorem \ref{thm:dilworth} to past-finite causets should be obvious: although the height of the causet will be infinite---and hence the partition will consist of infinitely many sets or `layers'---, it is still the case that every event has a height and that the resulting `height sets' will be antichains. The same technique cannot be applied in the case of causets which are neither past- nor future-finite,\footnote{An obvious variant of the same technique will work for future-finite causets.} though at least some of these infinite causets admit a foliation. Let me remind the reader that these result only guarantee the existence, but not of course the \emph{unique} existence of such a foliation; in fact, the foliations of a causet will in general be non-unique.

So there is a {\em problem of space} in causal set theory: there is a clear sense in which causets have no spatial structure. Given that the natural correlate of `space' in causal set theory are unstructured, inextendible antichains, can we nevertheless attribute some spatial structure to these antichains? Can we reconstruct spatial structure from a fuller structure? Could it be that perhaps a causal set induces some such structure on these antichains in a principled way? If so, then we would have completed an important step toward solving the problem of space in causal set theory. As it will turn out, this question cannot be addressed separately from the overarching problem of the emergence of spacetime from causets. So asking more generally, and in anticipation of the next chapter, is there a theoretically sound way of extracting geometrical information from the fundamental causets that could be used to relate them to the smooth spacetimes of GR? Before we get to the full geometry, and hence to how `manifoldlike' causets are, let us consider the more basic notions of dimension and topology, with a particular eye toward trying to identify any `spatial' structure in causets. 

\subsection{The dimensionality of a causal set}
\label{ssec:dimension}

The general idea of how to attribute a particular dimension to a causet, or of a partial order more generally, is to identify some property of the causet itself that is indicative of the dimensionality of the `smallest' manifold into which can be comfortably embedded (if it can be so embedded). Following \citet{bri97}, a standard definition of the dimension of a partially ordered set starts out from what is called the `coordinate order'. To illustrate that concept, consider the order structure imposed on the plane $\mathbb{R}^2$ by introducing an ordering relation $\leq$ on the elements of $\mathbb{R}^2$ such that $\langle x, y\rangle \leq \langle u, v\rangle$ just in case both $x\leq u$ and $y\leq v$, where we introduce Cartesian coordinates in the usual way such that a point $p\in\mathbb{R}^2$ is identified with an ordered pair of `coordinates', and where `$\leq$' is interpreted as the usual `less or equal' relation. $\leq$ induces a partial order on $\mathbb{R}^2$, as there are pairs of points in $\mathbb{R}^2$ such as $\langle 0,1\rangle$ and $\langle 1,0\rangle$ that are incomparable. The ordering can be straightforwardly generalized to yield what Brightwell (53) terms the {\em coordinate order} on $\mathbb{R}^n$ for any positive integer $n$. One can then define the dimension as follows (55):
\begin{defi}[Dimension of partial order]\label{def:dim}
The {\em dimension} dim$(\mathcal{P})$ of a partially ordered set $\mathcal{P} = \langle P, \preceq\rangle$ is the minimum integer $n$ such that $\mathcal{P}$ can be embedded in $\mathbb{R}^n$ with the coordinate order. 
\end{defi}
A common alternative, but equivalent, definition renders---of course---the same verdict. It defines the dimension of $\mathcal{P}=\langle P, \preceq\rangle$ as the smallest number of total ordering relations on $P$ whose intersection is $\preceq$, i.e., $\forall x,y \in P, x\preceq y$ if and only if $x$ is below $y$ in each of the total orders. This definition goes back to \citet{dusmil41} and is called `combinatorial dimension' in \citet[26]{mey88}, and sometimes `order dimension' elsewhere. 

What is the minimum condition on a mapping from $P$ to $\mathbb{R}^n$ that it qualifies as an `embedding'? In the general case at hand, the only substantive rule is that the embedding is `order-preserving', i.e., it is a mapping $f$ from $P$ to $\mathbb{R}^n$ that preserves the ordering in that $\forall p, q\in P, p\preceq$ q if and only if $f(p) \leq f(q)$, where `$\leq$' is the coordinate order as introduced above. 

Applied to the case at hand, since any two elements are unrelated in an antichain, its dimension must be equal to its cardinality: an antichain of two elements requires an embedding into $\mathbb{R}^2$ such that both elements can be mapped in a such a way that they are not related by the coordinate order, an antichain of three elements needs an embedding into $\mathbb{R}^3$, etc. 

What about the dimension for partial orders more generally? \citet{dusmil41} not only show that the dimension is well defined for every partially ordered set, but also prove that the dimension of a partially ordered set of size $n$ is finite if $n$ is finite and no greater than $n$ if $n$ is transfinite (Theorem 2.33). \citet[81]{hir51} strengthens this to the statement that the dimension of a partially ordered set is no greater than its size $n$, for $n$ finite or transfinite. He also shows (Theorem 5.1) the stronger claim that the dimension of a partially ordered set increases by at most 1 with the addition of a single element to the partially ordered set. 

While Definition \ref{def:dim} captures the intuition behind what it is to be a `dimension', and these results establish that `dimension' is well defined and has an upper bound, the notions suffer from three major limitations that make it unusable for present purposes. First, even though the dimension is well-defined, it is generally hard to compute it: for $n\geq 3$, the computation of whether a given partially ordered set has dimension $n$ is an \textsf{NP}-complete problem \citep[cf.][]{feleal14}. Second, it should be emphasized just how weak the established bounds are: for a universe the size of something like $10^{245}$ in Planck units, and hence with as many causet elements, we would like to strengthen the assertion that the fundamental causet has a dimension of at most $10^{245}$ by about 244 to 245 orders!

Third, and most importantly, Definition \ref{def:dim} needs to be modified to suit the present context: here we are not concerned with embedding partially ordered sets into $\mathbb{R}^n$, but rather with embedding causets into {\em relativistic spacetimes}. Thus, while the first two limitations may be of a rather technical nature, this third point cuts directly to the heart of what we are interested in in this book: the emergence of spacetime from structures in quantum gravity. Therefore, the connection between spacetimes and causets must be imbued with physical salience. Without it, there is no hope to advance the present project. For the purposes at hand, it is much more sensible to define the dimension of a causet in relation to relativistic spacetimes. We start by defining the relevant kind of embedding:
\begin{defi}[Embedding]\label{def:embed}
An {\em embedding} of a causet $\langle C, \preceq\rangle$ into a relativistic spacetime $\langle\mathcal{M}, g_{ab}\rangle$ is a injective map $f: C\rightarrow \mathcal{M}$ that preserves the causal structure, i.e., $\forall x, y\in C, x\preceq y \Leftrightarrow f(x) \in J^-(f(y))$.
\end{defi}
A causet that is embeddable, i.e., affords an embedding into a relativistic spacetime, could in this sense be regarded as a discrete approximation to a this spacetime---their causal structures are consistent. Of course, systematically speaking, the fundamentality of the causet demands that this relationship runs in the other direction: relativistic spacetimes, at least to the extent to which they present physically reasonable models of the large-scale structure of the universe are low-energy approximations to an underlying fundamental causet. Given Malament's Theorem \ref{thm:mala} it is evident, furthermore, that an otherwise unrestricted embedding of a causet into a spacetime cannot fully recover all salient features of the spacetime; at most, it approximately determines the conformal structure of the spacetime. Additional assumptions will be necessary to fix the conformal factor, although the cardinality naturally offers to regard the number of elements as a measure of the size. We will return to this problem in the next chapter; for the analysis of dimensionality it suffices to consider any embeddable causets. 

Although quantum gravitists are ultimately interested in the regime of strong gravitational fields, the embeddability of causets into (subspaces of) Minkowski spacetime provides an important test case. It is natural, then, to follow the seminal \cite{mey88} and to study the dimension of causets in terms of their embeddability into Minkowski spacetime. Meyer defines the `Minkowski dimension' as follows:
\begin{defi}[Minkowski dimension]
``The {\em Minkowski dimension} of a causal set is the dimension of the lowest dimensional Minkowski space into which it can be embedded (not necessarily faithfully).'' \citep[16f]{mey88}
\end{defi}
\citet{mey88,mey93} proves that the Minkowski dimension of a partial order is identical to its dimension as defined in Definition \ref{def:dim} in dimension two, but not in higher dimensions. It is natural to ask whether we can similarly find least upper bounds for dimensions higher than two. It turns out that it is not the case that a partially ordered set of size $n$ has a Minkowski dimension of at most $n$, as there are some finite ones that cannot be embedded in any finite-dimensional Minkowski spacetime \citep[cf.][]{feleal99}. In the absence of analytical results with much traction, different methods to estimate the Minkowksi dimension of causets have been developed, with some encouraging numerical results particularly for high-density `sprinklings'.\footnote{Most prominent among them are the `Myrheim-Meyer dimension' \citep{mey88} and the `midpoint-scaling dimension \citep{bom87}.} 

To return to the problem of space, it should be noted that all the work on the dimension of partially ordered sets assumes a non-trivial relational structure on the set and, in the case of causets, considers entire causets, and not just some unstructured subsets. It is not obvious what the Minkowski dimension of the antichains that are supposed to represent `space' at a `time' is, given that these were supposed to be interpreted purely spatially and Minkowski spacetime is defined by its causal structure, and thus is spatio\emph{temporal}. Although the outcome will arguably not miss its mark by as much as the standard dimension for partially ordered sets, these difficulties illustrate just how unlike space an antichain is. If space is to be found in causal sets, it will require more structure.

\subsection{The spatial topology of causal sets}
\label{ssec:topology}

This brings us to the next to final task for this chapter: is it possible to induce a topology on the `spatial' antichains which will endow them with a structure at least starting to look more `space-like'? Given that there is no structure in an antichain, we cannot hope to extract topological structure from it. It turns out that there is a way to {\em induce} a topology on the `spatial' antichains of a causet, but it requires much or all of the structure of the entire causet.

A {\em topology} on a set $X$ is a set $\mathcal{T}$ of subsets of $X$---the `open sets'---which satisfies the following conditions:
\begin{enumerate}
\item $\emptyset \in \mathcal{T}$ and $X\in \mathcal{T}$;
\item the union of a collection of sets in $\mathcal{T}$ is in $\mathcal{T}$;
\item the intersection of any two (and hence any finite number of) sets in $\mathcal{T}$ is in $\mathcal{T}$. 
\end{enumerate}
It turns out that for a finite $X$, there exists a one-to-one correspondence between topologies on $X$ and preorders on $X$, where a {\em preorder} is a binary relation that is reflexive and transitive. Apart from the fact that a spatial antichain may not be finite, this theorem does not help us to obtain a natural topology on the spatial antichain simply because there is no fundamental physical relation at all that obtains between the elements of the antichain, and a fortiori no preorder. Case closed?

Let's not jump to conclusions; of course, one can easily {\em impose} topologies on an altogether unstructured set. For any such set $X$, just set $\mathcal{T} = \{ X, \emptyset\}$---the so-called {\em indiscrete topology} on $X$. But such a coarse topology will not further our goal in identifying some useful geometrical structure on a spatial antichain. At the other end of the spectrum, we find the {\em discrete topology} $\mathcal{T} = \mathcal{P}(X)$, the set of all subsets of $X$ and hence the `finest' possible topology on $X$. But two cautionary remarks expose the very limited appeal of the discrete topology for our purposes.

First, and as a first indication of the need to go beyond a `spatial' antichain, imposing the indiscrete topology seems to presuppose a distinction between the set and the empty set. In other words, a presupposition of the indiscrete topology is that $X$ is non-empty. That seems harmless enough. But by the same token, the discrete topology seems to assume not only that there are elements in $X$, but that there are {\em numerically distinct} elements. And that is decidedly less harmless. If basal events in causal set theory are indeed featureless and $\preceq$ is the only physical relation at the fundamental level, then any two events with the same relational profile, i.e., a non-Hegelian pair, cannot be physically distinguished \citep[cf.][]{wut12,wutcal15}. In an antichain, of course, elements cannot be so distinguished. This raises two challenges. The first is to metaphysically underwrite a difference between a singleton set and a set containing a plurality of events. If we accept primitive plurality even in cases the elements cannot be physically distinguished---as we may have to do for independent reasons \citep{wut09}---, then that problem can be circumvented (and the presupposition in the indiscrete case secured). But there remains a second, and harder, challenge. Merely to be able to assert that the antichain consists of $n$ elements does not suffice to distinguish the various subsets of the antichain; even just for $n=2$, there would have to be two distinct singleton sets in the discrete topology. But with nothing to distinguish the elements they contain, nothing can mark their distinctness. Of course we can assert that there must be {\em two} such singleton sets, but for a topology to be able to structure the set in a meaningful way, it seems necessary to be able to tell one of them apart from the other. This could be accomplished by insisting that the basal events of causal set theory command some haecceitistic identity; however, haecceities offend against the natural structuralist interpretation of causal set theory \citep{wut12}. 

It should be noted that if the basal elements in the antichain $X$ enjoy numerical distinctness, then we can in fact define a distance function $d$ such that $\langle X, d\rangle$ is a metric space. For instance, for any $x, y \in X$, define
\begin{equation}
d(x, y) = \left\{
  \begin{array}{l l}
    0, & \quad \text{if $x=y$,}\\
    1, & \quad \text{if $x\neq y$.}
  \end{array} \right.
\end{equation}
This function straightforwardly satisfies the standard conditions demanded of a distance function (i.e., nonnegativity, nondegeneracy, symmetry, and the triangle inequality), and $\langle X, d\rangle$ thus qualifies as a metric space (called a {\em discrete metric space}). Clearly, however, this $d$ will not not give rise to any `space' that resembles physical space as found, e.g., in GR. And it remains opaque how a structure with such a distance function could be meaningfully embedded in a spacetime of low dimensionality. Furthermore, imposing topologies and metrics on a `spatial' antichain seems to require a metaphysics of basal events that is anathema to the spirit of causal set theory.

Second, then, even granting this possibility of inflicting a topology and indeed a metric of our `spatial' antichains, no physically useful structure can be extracted from them. A topology on a `spatial' slice should underwrite nearness relations. In order to successfully do that, it shouldn't be too coarse---as was the indiscrete topology---for then it would seem as if no events are `nearby' one another. However, it also shouldn't be too fine---as was the discrete topology---for otherwise {\em all} events are near all other events. Either way, there are no physically useful nearness relations that appropriately discriminate some events to be nearer than others. So we will want a `goldilocks' topology that finds the sweet spot in fineness. Furthermore, the topology of `spatial' antichains should cohere with, and in fact give rise to, the nearness relations as we find them in spatial slices of the emerging relativistic spacetime. This means that it should be appropriately grounded in the fundamental physics in place. Thus, our best bet to impose some structure in general, and topological structure in particular, on `spatial' slices of causal sets is to start out from the one physical relation present in a causal set---$\preceq$. 

The imposition of a physically perspicuous topology based on $\preceq$ can be accomplished by considering how an inextendible antichain is embedded in the causet. \citet*{Major2006,Major2007} offer just such a way of doing that. Their construction proceeds by `thickening' the antichain and exploiting the causal structure gained by this `thickening'. The idea is, roughly, as follows. For any subset $X$ of the domain $C$ of a causal set, define $F(X) := \{ x\in C |\; \exists y\in X, y\preceq x\}$---the causal future of $X$---and $P(X) := \{x\in C |\; \exists y\in X, x\preceq y\}$---the causal past of $X$. Note that given the reflexive convention chosen for $\preceq$, $x\in F(x)$ and $x\in P(x)$ for any $x$ in $C$. Next, define the {\em $n$-future thickening} of an (inextendible) antichain $A\subset C$ as
\begin{equation}
A_n^+ := \{x\in C |\; x\in F(A) \;\mbox{and}\; |P(x)\setminus P(A)| \leq n\},
\end{equation}
where $|X |$ denotes the cardinality of set $X$, and, mutatis mutandis, the {\em $n$-past thickening} $A_n^{-}$ of $A$. Note that---again as a consequence of the reflexive convention---$A_0^+ = A_0^{-} = A$ for any antichain $A$. $A_1^+$ and $A_1^{-}$ will contain the antichain together with all the immediate successors and predecessors of events in the antichain, respectively, etc. See figure \ref{fig:topology1} for an example of $A_1^+$.
\begin{figure}[ht]
\centering
\epsfig{figure=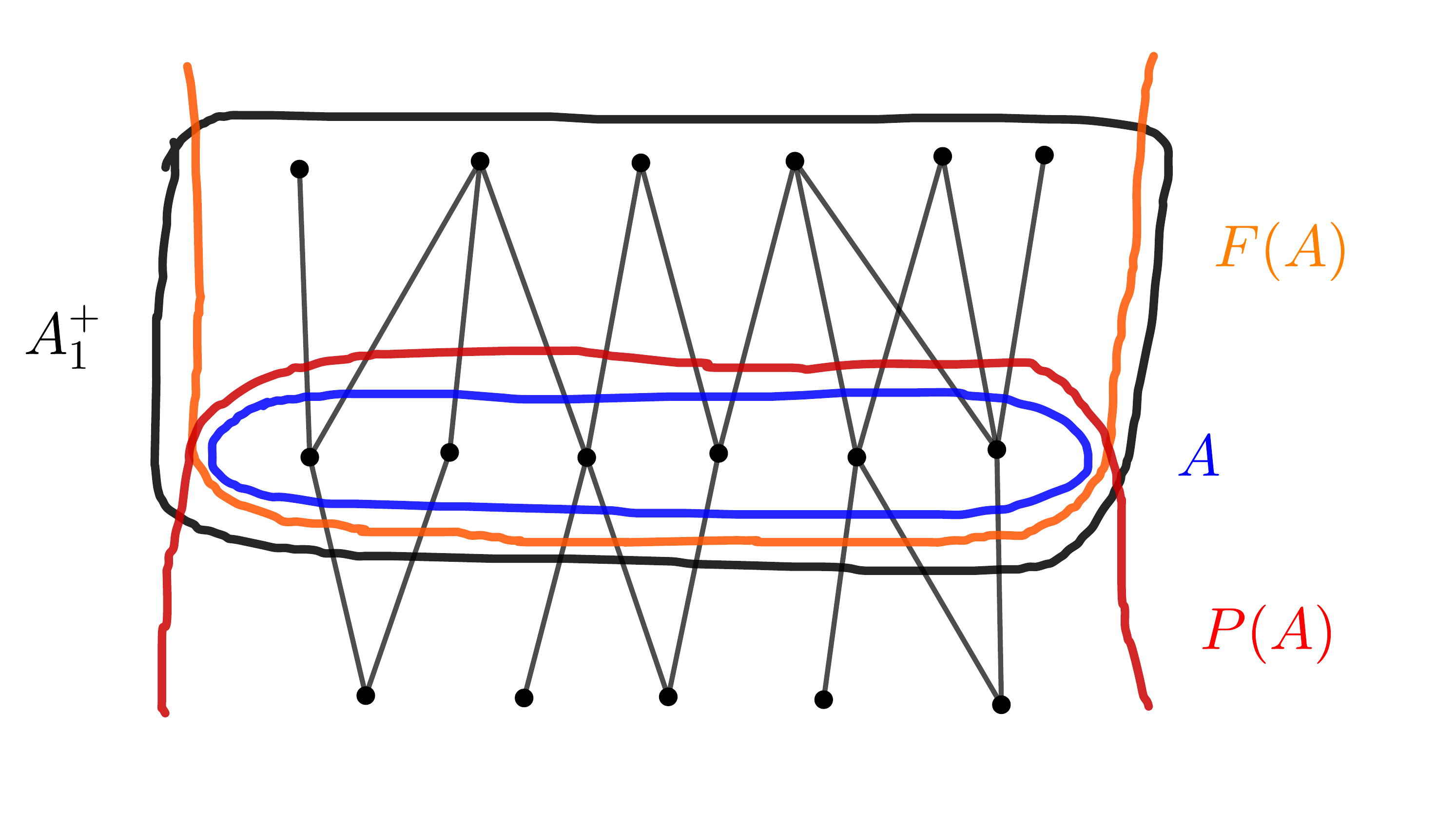,width=0.75\linewidth}
\caption{\label{fig:topology1} A 1-future thickening $A^+_1$ of an antichain $A$.}
\end{figure}
Finally, define an {\em $n$-thickening} $A_n$ of $A$ as $A_n^+ \cup A_n^{-}$. 

Next, identify the maximal or `future-most' elements $m_i$ of $A_n^+$ and form the sets $P_i := \{ m_i |\; P(m_i) \cap F(A)\}$ of `past lightcones' of the maximal elements truncated at the `spatial' antichain $A$. Then $\mathcal{P} := \{P_i\}$ is a {\em covering} of $A_n^+$, i.e., $\cup_i P_i = A_n^+$. From these, we can construct the {\em shadow sets} $A_i := P_i \cap A$. See figure \ref{fig:topology2} for an example of a shadow set.
\begin{figure}[ht]
\centering
\epsfig{figure=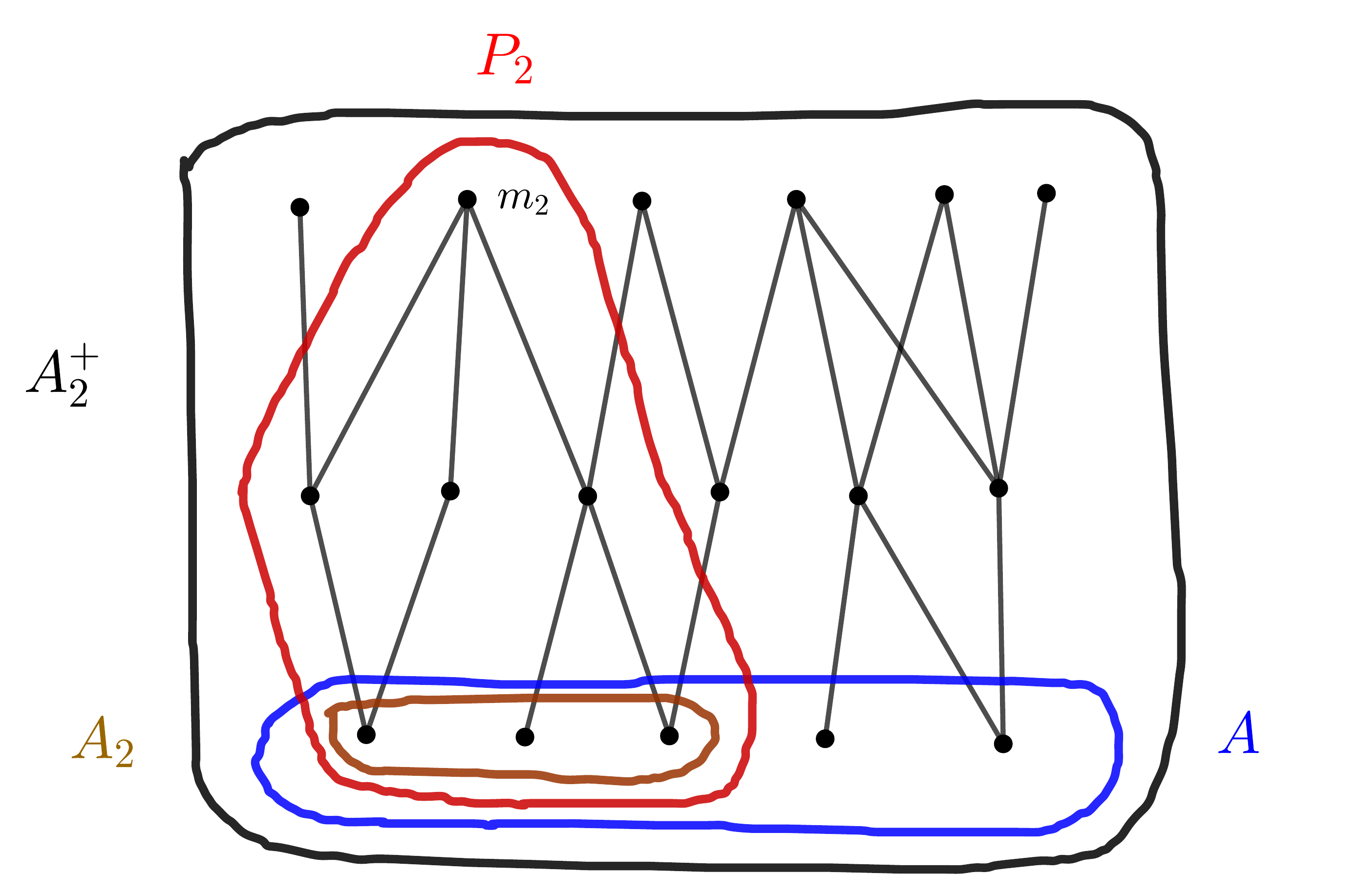,width=0.7\linewidth}
\caption{\label{fig:topology1} A maximal element $m_2$ of a 2-future thickening $A_2^+$ of an antichain $A$ and the resulting shadow set $A_2$.}
\end{figure}
We find that $\mathcal{A} := \{A_i\}$ provides a covering of $A$ and can be used as a vantage point to specify a topology on $A$. $\mathcal{A}$ itself will not, in general, be a topology because it may not satisfy any of the three conditions on topologies. \citet{Major2007} use the shadow states to construct topological spaces called `nerve simplicial complexes' and show that natural continuum analogues of these topological spaces are homotopic to the globally hyperbolic spacetimes in which they are defined. This is a hopeful sign that the topologies constructed in this manner are indeed the physically salient ones, assuming, of course, that the topologies of these globally hyperbolic spacetimes are. Furthermore, their results contribute toward establishing a relevant form of the so-called {\em hauptvermutung} and hence to the emergence of spacetime from causal sets. We will turn to these topics in the next chapter. 

It should be noted that the topologies that will arise from just the maximal elements of an $n$-future thickening for some one particular $n$ are generally going to be too coarse to be fully satisfactory. However, any such topology can be refined by adding shadow sets arising from thickenings with different $n$, and from the analogous construction based on past thickenings and `minimal' elements. The finest topology obtainable on a `spatial' antichain $A$ will result from thickening in both causal directions and letting $n$ range over all values $0,..., N$, for an $N$ such that $A_N^+\cup A_N^{-} = C$. The finest topology for an antichain will thus be obtained from considering the full structure of the causet it inhabits. It should thus be noted that the spatial structure---in this case the topology of `spatial' slices---asymmetrically depends on the fundamental causal structure. 

The constructions used to introduce topologies on `spatial' slices can be used to introduce something akin a covariant sum-over-histories approach to dynamics, as articulated by \citet{Major2006}. Informally, the picture resembles a three-layered cake with the middle bulk sandwiched between an `initial' state of the universe consisting of the minimal elements of some $A_n^{-}$ and a `final' state constituted by the maximal elements of an $A_m^+$. The cake would then represent a `spatial' slice that evolves from some `initial' state to some `final' state. Interestingly, it is possible to define physically meaningful `transition amplitudes' from the `initial' to the `final' state as measures over the set of completed causets containing the fixed `top' and `bottom' layer of the relevant `cake', but with generally differing `interpolators' between them. 

\subsection{Finding distances}
\label{ssec:distance}

A next step would be to recover metric structure from the causet. For `timelike' distances, i.e., something like durations, there is a widely accepted recipe to obtain a straightforward proxy of continuum timelike distances. In analogy to relativistic spacetimes, one defines a geodesic in causal set theory as follows:
\begin{defi}[Geodesic]
A {\em geodesic} between two elements $x\preceq y \in C$ is the longest chain $\gamma$ from $x$ to $y$, i.e., the chain of the largest cardinality with past endpoint $x$ and future endpoint $y$. If a geodesic $\gamma$ has cardinality $n+1$, i.e., $|\gamma | = n+1$, then its {\em length} is $n$. 
\end{defi}
In general, a geodesic between two elements is not unique. That geodesics thus defined offer a natural analogue of their continuous cousins was first conjectured by \citet{myr78}. \citet{Brightwell1991} showed that for causal sets that are approximated (see next chapter) by a flat spacetime interval, for sufficiently large geodesics, the length of geodesics in those causets rapidly converges to (a multiple) of the proper time elapsed between the images of the endpoints in the spacetime. Though analytic results are only available for low dimensions and flat spacetimes, numerical studies \citep{ilieal06} suggest that the convergence holds also in dimensions up to 4 and in some curved spacetimes.\footnote{Cf.\ also \citet[\S 1.3.2]{ridwal09a}.} 

\begin{figure}
\centering
\epsfig{figure=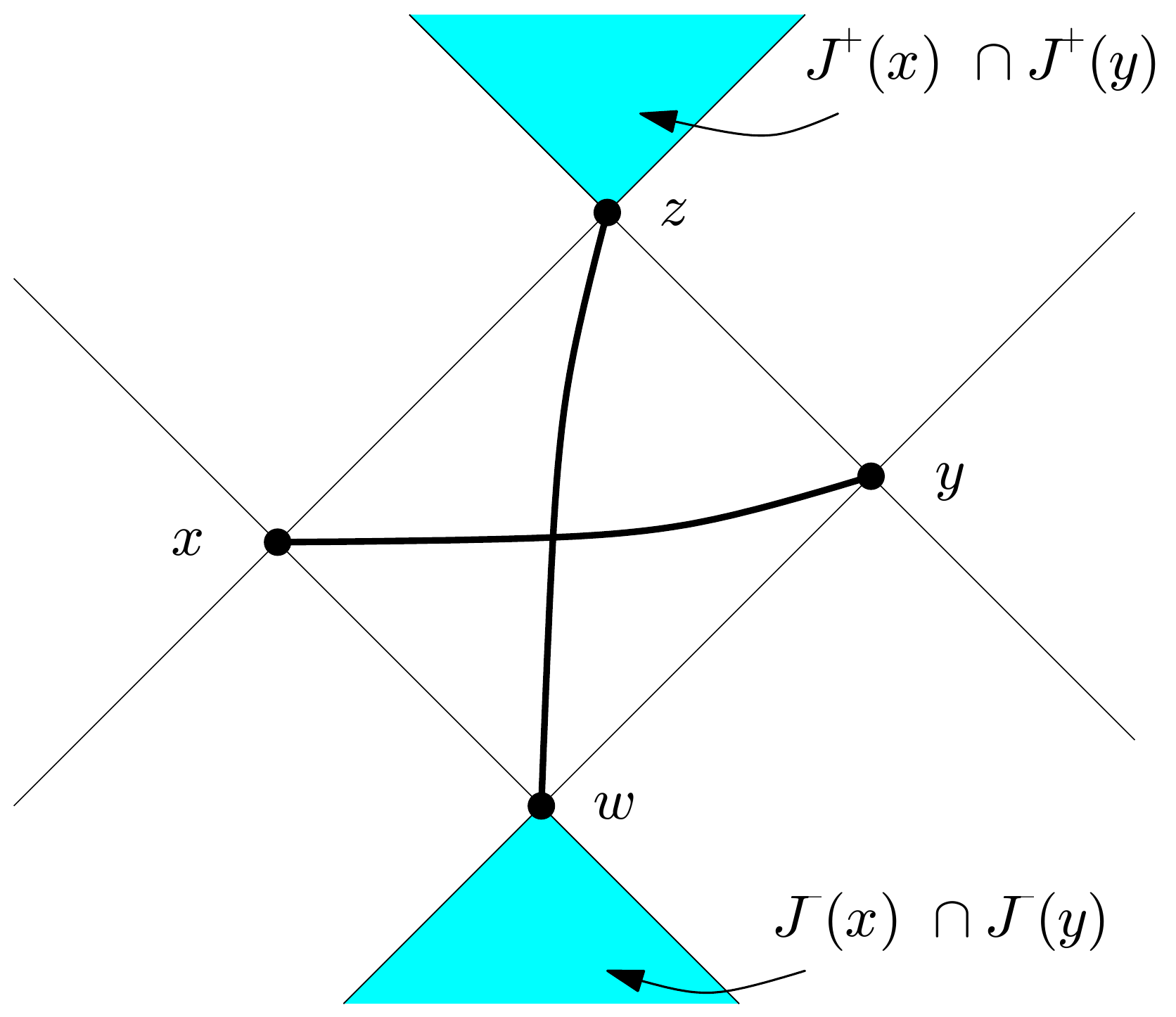,width=0.5\linewidth}
\caption{\label{fig:spatialdistance} A first stab at determining the spatial distance.}
\end{figure}

Let us return to the topic of this section, `space'. Spatial distances can also be introduced, albeit much more vicariously, but they too rely on the structure of the causet not contained in the antichain. \citet{ridwal09a} have offered the most penetrating analysis of spatial distance to date. An obvious first attempt (cf.\ Figure \ref{fig:spatialdistance}) starts out from the continuum case and tries to generalize that to the discrete one. It defines the spatial distance between two spacelike separated events $x$ and $y$ in terms of an appropriate timelike distance, since we already know (from the previous two paragraphs) how to determine the timelike distance between two events in a causet. In the continuum case, let $w$  be an event in the common past $J^- (x) \cap J^- (y)$ and $z$ one in the common future $J^+ (x) \cap J^+ (y)$ such that the timelike distance between them is shortest for all such pairs. In other words, $w$ is a maximal element of the common past and $z$ is a minimal element of the common future (see Figure \ref{fig:spatialdistance}). 
Then the spatial distance between $x$ and $y$ is equal to the timelike distance between $w$ and $z$ in the continuum.  For two-dimensional Minkowski spacetime, the pair $w$ and $z$ is unique. In higher dimensions, however, this is not the case: in $d$-dimensional Minkowski spacetime, there is a $(d-2)$-dimensional submanifold of pairs of events that minimize the timelike distance \citep[7]{ridwal09a}. 

This non-uniqueness thwarts the application of this simple recipe to the discrete case of causal sets that can only be embedded into higher-dimensional Minkowski spacetimes. Details of the embedding and the `sprinkling' will be discussed in the next chapter, but the problem turns out to be that given the required kind of `sprinkling' the events in the causet into Minkowski spacetime, there is a finite probability for each of the infinitely many minimizing pairs that the Alexandrov interval they encompass does not contain any of the images of events in the causet and is thus empty of `sprinkled' events. Consequently, there will always be some pair $\langle w, z\rangle$ such that $w\preceq x\preceq z$ and $w\preceq y \preceq z$ are the longest chains between $w$ and $z$ (since $x\not\preceq y$), both of which are of length 2. Hence, the timelike distance between $w$ and $z$, and thus the spatial distance between $x$ and $y$, will always be 2. 

Although this would technically give us a distance function, it can hardly be considered physically adequate. \citet{ridwal09a} analyze a number of more involved approaches to extracting non-degenerate spatial distances from the fundamental structure of the causet. Their results are limited, but promising. As far as we can tell, they all resort to considering substantively larger parts of causets than merely unstructured antichains. Once again, this reflects the fundamentality of the causal structure over any spatial or temporal structure. 

\subsection{Wrapping up}
\label{ssec:wrapping}

Relating causets to spacetimes via their `spatial' and `temporal' parts has thus thoroughly failed, in more radical ways even than in GR. In the next chapter, we will consider how they might be related {\em in toto}, as wholes. For now, we can only state just how different causets are from spacetimes, lest we are inclined to see the causal relation at the core of the models of causal set theory as ultimately spatiotemporal, as it arguably is in GR. The geometric structure that we would normally attribute to space in particular, and that is certainly available in GR, such as topological, affine, differential, and metric structure is only very indirectly recoverable from the structure of causets, if at all. It is simply not built in at the fundamental level. In this sense, causal set theory offers a view of our world that is not ultimately spatiotemporal. In the next chapter, we will analyze how relativistic spacetime could emerge from fundamental causal sets and how spacetime functionalism will helpfully delimit the to-do list to establish this emergence. In this analysis, we will identify the work sketched in \S\ref{sec:space} as functionalist. In order for emergence to succeed, dynamical laws beyond the kinematics studied in this chapter will turn out to be called for. We will complete our discussion of causal set theory with two philosophically fruitful points that arise in connection to the dynamical aspects of the theory: the possibility of relativistic becoming and a form of non-locality that has nothing to do with quantum physics.

\bibliographystyle{plainnat}
\bibliography{biblio}

\end{document}